\documentclass{article}

\usepackage{arxiv}

\usepackage[utf8]{inputenc} 
\usepackage[T1]{fontenc}    
\usepackage{hyperref}       
\usepackage{url}            
\usepackage{booktabs}       
\usepackage{amsfonts}       
\usepackage{nicefrac}       
\usepackage{microtype}      
\usepackage{lipsum}
\usepackage{graphicx}
\usepackage[square,sort,comma,numbers]{natbib}

\hypersetup{%
  colorlinks=true,
  linkcolor=blue,
  citecolor=blue,
  urlcolor=blue
}

\title{Submillimetre Transient Science in the Next Decade}

\author{
  Steve Mairs\thanks{\texttt{\url{s.mairs@eaobservatory.org}}}$\:\:^{1}$~~$\bullet$~~ 
  Gregory Herczeg$^{2}$~~$\bullet$~~ 
  Doug Johnstone$^{3}$~~$\bullet$~~
  Jeong-Eun Lee$^{4}$~~$\bullet$~~
  Simon Coud\'e$^{5}$\\
 \textbf{Alexandra J. Tetarenko}$^{1}$~~$\bullet$~~
 \textbf{Jenny Hatchell}$^{6}$~~$\bullet$~~
 \textbf{Aleks Scholz}$^{7}$~~$\bullet$~~
 \textbf{Bhavana Lalchand}$^{8}$~~$\bullet$~~
 \textbf{Wen-Ping Chen}$^{8}$\\
 \textbf{Carlos Contreras Pe\~{n}a}$^{6}$~~$\bullet$~~
 \textbf{Tim Naylor}$^{6}$~~$\bullet$~~
 \textbf{Kevin Lacaille}$^{9}$~~$\bullet$~~
 \textbf{Peter Scicluna}$^{10}$\\\\
   \textit{$^{1}$East Asian Observatory (JCMT) 660 N. A`oh\={o}k\={u} Place, Hilo, Hawai`i, USA, 96720}\\
   \textit{$^{2}$Kavli Institute for Astronomy \& Astrophysics, Peking University, Beijing, China}\\
   \textit{$^{3}$NRC Herzberg Astronomy and Astrophysics, 5071 West Saanich Rd, Victoria, BC, V9E 2E7, Canada}\\
   \textit{$^{4}$School of Space Research, Kyung Hee University, Giheung-gu Yongin-shi, Gyunggi-do, 17104, Korea}\\
   \textit{$^{5}$SOFIA Science Center, USRA, NASA Ames Research Center, Moffett Field, CA, USA, 94035}\\
   \textit{$^{6}$Physics and Astronomy, University of Exeter, Stocker Road, Exeter EX4 4QL, UK}\\
   \textit{$^{7}$SUPA, School of Physics \& Astronomy, North Haugh, St. Andrews, KY16 9SS, UK}\\
   \textit{$^{8}$Graduate Institute of Astronomy, National Central University, 300 Zhongda Rd. Zhongli, Taoyuan 32001, Taiwan}\\
   \textit{$^{9}$Department of Physics and Atmospheric Science, Dalhousie University, Halifax, NS, B3H 4R2, Canada}\\
   \textit{$^{10}$Academia Sinica Institute of Astronomy and Astrophysics, AS/NTU Astronomy-Mathematics Building,}\\\textit{No 1. Sec. 4 Roosevelt Rd, Taipei, Taiwan}
}

\begin{document}
\maketitle

\begin{abstract}

This white paper gives a brief summary of the time domain science 
that has been performed with the JCMT in recent years and highlights 
the opportunities for continuing work in this field over the next 
decade. The main focus of this document is the JCMT Transient Survey, a
large program initiated in 2015 to measure the frequency and amplitude 
of variability events associated with protostars in nearby star-forming
regions. After summarising the major accomplishments so far, an outline
is given for extensions to the current survey, featuring a discussion
on what will be possible with the new \mbox{850 $\mu$m} camera that is
expected to be installed in late 2022. We also discuss possible applications of submillimetre monitoring to active galactic nuclei, X-ray binaries, asymptotic giant branch stars, and flare stars.
\end{abstract}



\section{Introduction}
\label{sec:intro}
 
The initial phase of the growth of a protostar occurs steadily, driven by the gravitational infall of material in the surrounding, dusty envelope ($\sim1000-10000$ AU). 
A protoplanetary disk ($\sim 0.1-100$ AU) forms early in this process \citep{Jorgensen2008}. Once formed, the disk channels most of the accreting material from the envelope to the protostar via a loss of angular momentum, likely due to viscous interactions and MHD instabilities. Finally, 
the mass is funnelled onto the protostar by the stellar magnetic field, which disrupts the disk at scales typically of order a few stellar radii 
\citep[For a review on accretion processes, see][]{Hartmann2016}. Once a disk forms, the accretion rate is expected to be 
variable due to instabilities in both the inner and outer disk \citep[see review by][]{Armitage2015}. This variability in
the rate of accretion has far-reaching implications for many of the most important aspects of star formation, including estimating
protostellar lifetimes \citep{Offner2011}, reconciling a decades-old discrepancy between theoretical and
observed brightnesses of young stars known as the ``Luminosity Problem'' \citep{Kenyon1990,Evans2009,Dunham2015}, and describing the 
physical structure of the circumstellar disk that will go on to form planets \citep{Bae2014,Vorobyov2015}. 


When a star accretes, most of the gravitational energy from the infall radius is converted into radiation \citep{stahler88}.  Any change in the accretion rate is expected to lead to a similar change in the total luminosity of the protostar.  The 
amplitudes and frequency of variability 
events associated with the changing accretion rate can inform us about the dominant 
physical drivers of unsteady mass accretion over time, but are virtually unconstrained in the literature. For example, bursts that last decades or even centuries suggest gravitational instabilities or 
processes in the outer disk, whereas short-term variability likely traces inner-disk/magnetic effects. 

Many wide-field and all-sky monitoring optical and near-IR surveys, including VVV \citep{VVV}, ASAS-SN \citep{ASASSN}, Gaia alerts\footnote{\url{http://gsaweb.ast.cam.ac.uk/alerts/home}}, Kepler K2 \citep{K2}, WASP \citep{WASP}, and the Zwicky Transient Facility (ZTF, \cite{ZTF}; previously PTF) have made significant contributions to young star variability.  Other monitoring campaigns, such as YSOVar \citep{YSOVAR}, have been dedicated explicitly to monitoring star-forming regions to evaluate variability.  These surveys have established the ubiquity of short-wavelength variability in optically-bright young stars, which are already nearly fully formed, 
but cannot probe the earliest, dominant stages of protostellar growth.

Accretion variability in the youngest sources has been challenging to study directly.  While it has been inferred to be common based on population studies of bolometric temperatures and luminosities, such as the Spitzer cores-to-disks program and the Herschel HOPS survey \citep{Evans2009,fischer2017}, at these stages the central 
protostar is heavily extincted by the nascent, dusty envelope. This dust, however, rapidly heats or cools in response to changes in the luminosity of the central source \citep{Johnstone2013}.  The envelope, which is well-traced by submillimetre observations, will brighten as the temperature increases (after, for example, a protostellar outburst associated with accretion) and it will dim with decreasing temperature. The typical timescale of these changes in submillimetre flux is expected to be weeks to months \citep{Johnstone2013}. 
While the strongest signal 
from a protostellar outburst would be expected at mid to far-infrared wavelengths, there is a current lack 
of space telescopes available to carry out consistent, regular observations of star-forming regions. Therefore, in order to probe these critical 
stages of mass accretion in Young Stellar Objects (YSOs), we capitalise on longer wavelength data. 

In an effort to investigate the changing mass accretion rate of stars during their earliest stages of formation, the James Clerk Maxwell Telescope (JCMT) Transient Survey Large Program was initiated \citep{Herczeg2017}. This program was the first, dedicated survey to monitor deeply embedded YSOs at submillimetre wavelengths, opening a new field in time-domain astronomy. 
The Transient Survey employs the Submillimetre Common User Bolometer Array 2 (SCUBA-2) to observe 8 nearby ($<500\mathrm{\:pc}$) star-forming regions (circular fields of 0.5$^{\circ}$ diameter) at an approximately monthly cadence.  Due to the development of novel relative flux calibration techniques that have decreased the flux uncertainty by an unprecedented factor of $2-3$ \citep{Mairs2017Cal}, more than a half dozen protostellar variables have been confirmed, including the most luminous stellar flare ever recorded \citep{Mairs2019}.  The early results from this survey have prompted both theoretical and observational follow-up studies by international teams within the JCMT partnership.  In the following sections, we describe the current status and future prospects for the Transient Survey, and subsequently discuss other possible applications of time-monitoring in the submillimetre.




\section{Current Status of Submillimetre Transient Science}
\label{sec:current}

The JCMT Transient Survey is the first dedicated program to monitor the light curves of compact, submillimetre sources. The survey began in December 2015 and will continue in its current form through at least January 2020.  While submillimetre 
variability associated with young, deeply embedded YSOs has been found before for a few objects \citep[e.g.][]{Safron2015}, the construction and refining of a 
relative flux calibration pipeline that improves the flux uncertainty at the telescope by a factor of \mbox{$2$ to $3$} 
\citep{Mairs2017Cal} has made it possible to monitor 1643 sources, $\sim 50$ with an accuracy of $\sim 2$\% and the remaining sample for any large changes. 
This JCMT large program has conclusively shown that $\sim 10-20$\% of the 50 brightest sources vary in brightness over timescales 
of months to years. These long-term brightness changes are interpreted as the dusty envelopes' response to luminosity changes in embedded 
YSOs. 

\begin{figure*}
  \centering
  \includegraphics[width=0.45\textwidth]{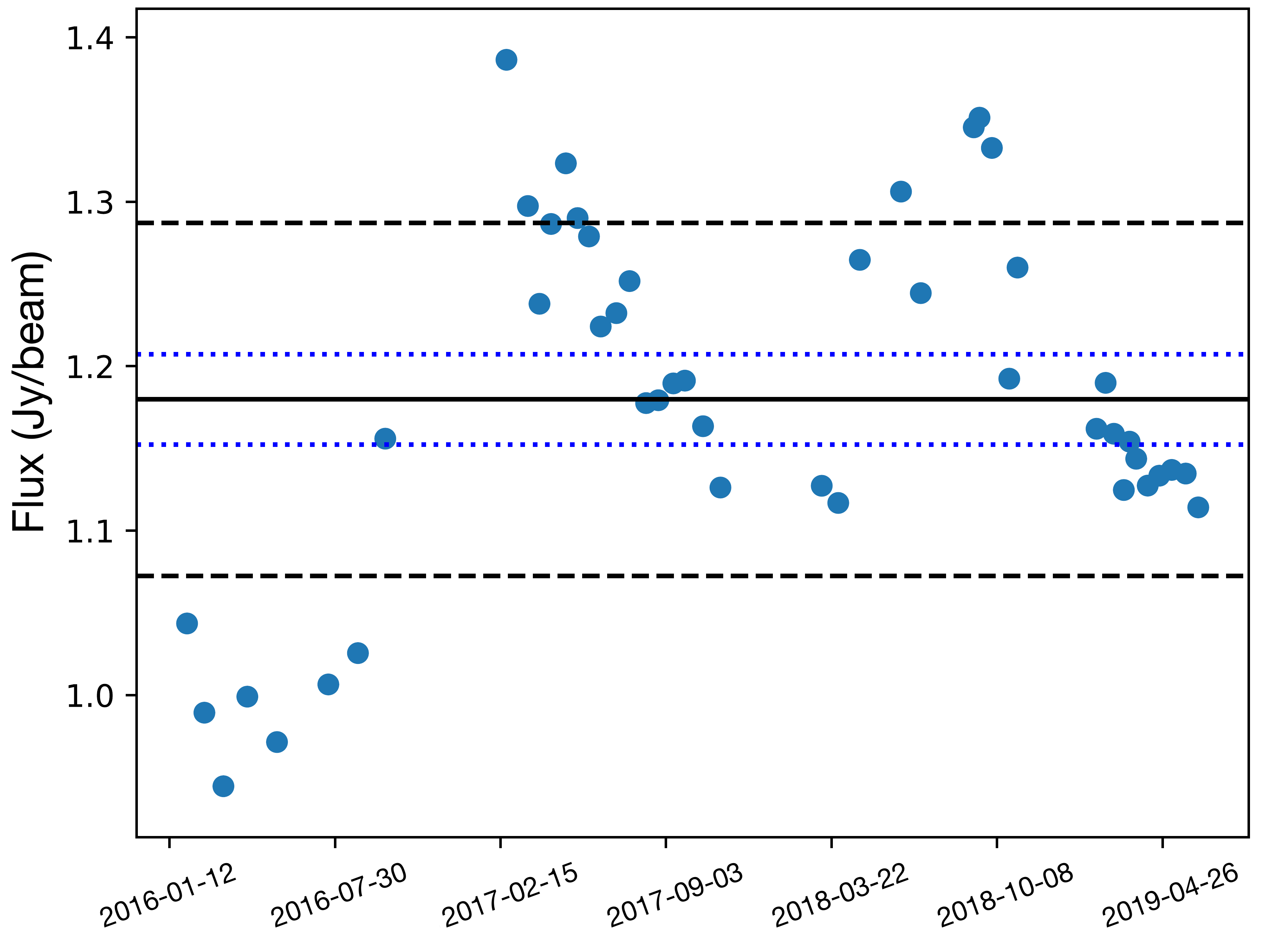}
  \includegraphics[width=0.45\textwidth]{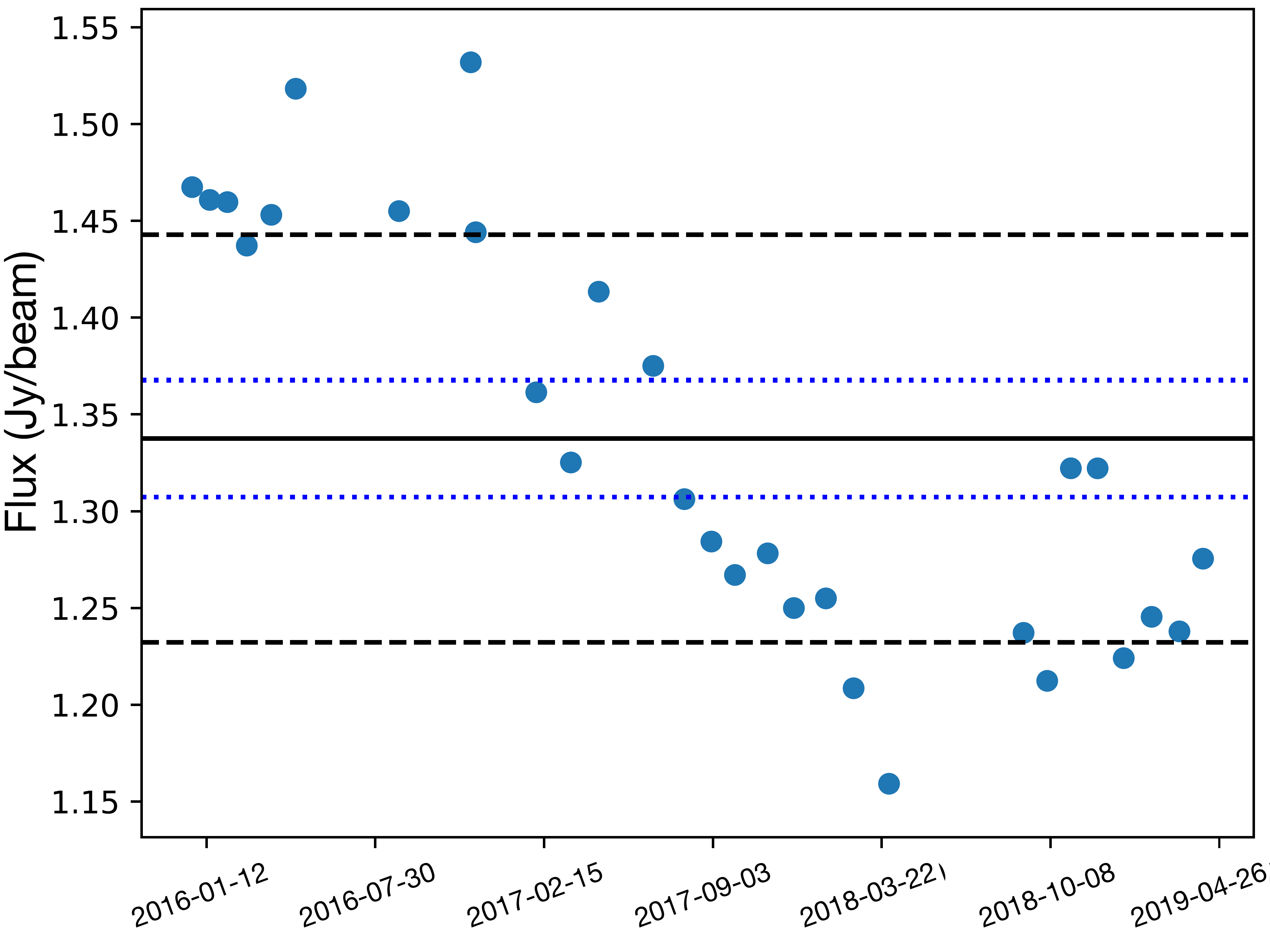}
  \caption{\textit{Left:} The \mbox{850 $\mu$m} light curve of \mbox{EC 53}. The variability period is \mbox{$\sim$543 days}. This variation is also seen at 450 microns, and at near-infrared wavelengths \protect\citep{Yoo2017}. The blue (dotted) lines indicate the expected light curve standard deviation in the absence of variability, while the black (dashed) lines indicate the measured light curve standard deviation. \textit{Right:} The \mbox{850 $\mu$m} light curve of \mbox{HOPS 358}. After a stable period of over one year, the flux steadily declined, then began increasing again. Additional data is necessary to determine if this is a periodic trend.}
  \label{fig:Variables}
\end{figure*}


The first prominent 
detection of an obvious variable in the survey was associated with the source \mbox{EC 53} \citep{Yoo2017}, also known as \mbox{V371 Ser} 
\citep{Hodapp2012}. The light curve shows a periodic trend of brightening and fading over a timescale of $\sim543\mathrm{\:days}$ (see left panel 
of Figure \ref{fig:Variables}). The periodicity is interpreted as accretion variability in the inner disk surrounding \mbox{EC 53}, perhaps excited by binary 
interactions. This variation and continual rise in brightness is seen at both 450 and \mbox{850 $\mu$m} and it is tightly correlated with near-infrared 
wavelengths \citep[J, H, and K bands; Tim Naylor, Watson Varricatt; private communications, ][]{Yoo2017}. The periodic nature of both the 
near-infrared and submillimeter wavelengths make the system a unique laboratory to study protostellar outbursts and how they inform the physics 
of the accretion disk. The short period indicates a timescale similar to EXor outbursts, but the NIR spectral features obtained by IGRINS suggest FUor-like characteristics. This indicates that the cooling timescale at the disk midplane must be longer than the  $\sim$1.5 year period, since the FUor-like NIR features can be explained by a hotter midplane than the surface of the disk.  There are currently ongoing investigations into the scaling relationships and spectral index of the source across these 
wavelengths.

EC 53 is, so far, the only known periodic submillimetre variable identified in the eight Transient Survey fields. Additional examples of long-term 
variability, however, have been discovered by performing several statistical tests on the 1643 identified \mbox{850 $\mu$m} sources across all 
observed regions \citep{Johnstone2018}. These tests are part of an automated pipeline that is triggered each time new data is obtained.  
In total, ten submillimetre variables have been confirmed within the Transient survey (see Table 7 of \citep{Johnstone2018}), while several additional candidates have been identified. \mbox{HOPS 
358} \citep{Furlan2016, Mairs2018Atel}, among the youngest and most deeply embedded 
YSOs in NGC 2068 (classified as a PACS Bright Red Source, \citep{Stutz2013}), has a strong brightness variation seen in Figure \ref{fig:Variables}. Further evidence of relatively long-term variability trends was found by identifying significant, robust changes 
in brightness (both brightening and fading) for 5 submillimetre sources observed 2-4 years apart by combining JCMT Transient 
Survey data with archival JCMT Gould Belt Survey \citep{Mairs2017GBS}, with further analysis of stochastic and 
secular variables in Lee et al.~(in Prep).

In addition to this long-term variability associated with accretion rate changes, the Transient Survey has also uncovered a non-thermal, short-term 
variability event signalling what may have been the most luminous stellar radio flare on record \citep{Mairs2019}. On 2016 November 26, a bright point source was 
detected in the direction of the T Tauri Binary system known as JW 566 \citep{Jones1988}. There has been no significant signal at this location 
during any of the other 26 Transient Survey observations, including data that was observed only 6 days previous to the flare. Upon further 
investigation, a light curve was constructed that showed the brightness of the source declining by 50\% in less than 30 minutes. The resulting 
brightness temperature suggests a non-thermal origin. Short-timescale, non-thermal variability similar to this has been noted before at millimetre 
and radio wavelengths \citep{Bower2003, Massi2006, Salter2008, Forbrich2008} but this is the first detection in the submillimetre regime. The flare 
is interpreted as a magnetic reconnection event, releasing (gyro-)synchrotron radiation. Additional observations of short-term variability associated 
with T-Tauri stars or younger YSOs will help determine the amplitudes and frequencies of these events. This will be an important window into the 
dominant physics governing material in the scale of the inner accretion disk to the stellar surface. High resolution spectral follow-up studies are 
currently under preparation. New methods are also under development to search for additional faint, short-term variability events in each 
observed epoch. These results will be summarised in Lalchand et al. (In Prep).

\begin{figure*}
  \centering
  \includegraphics[width=0.45\textwidth]{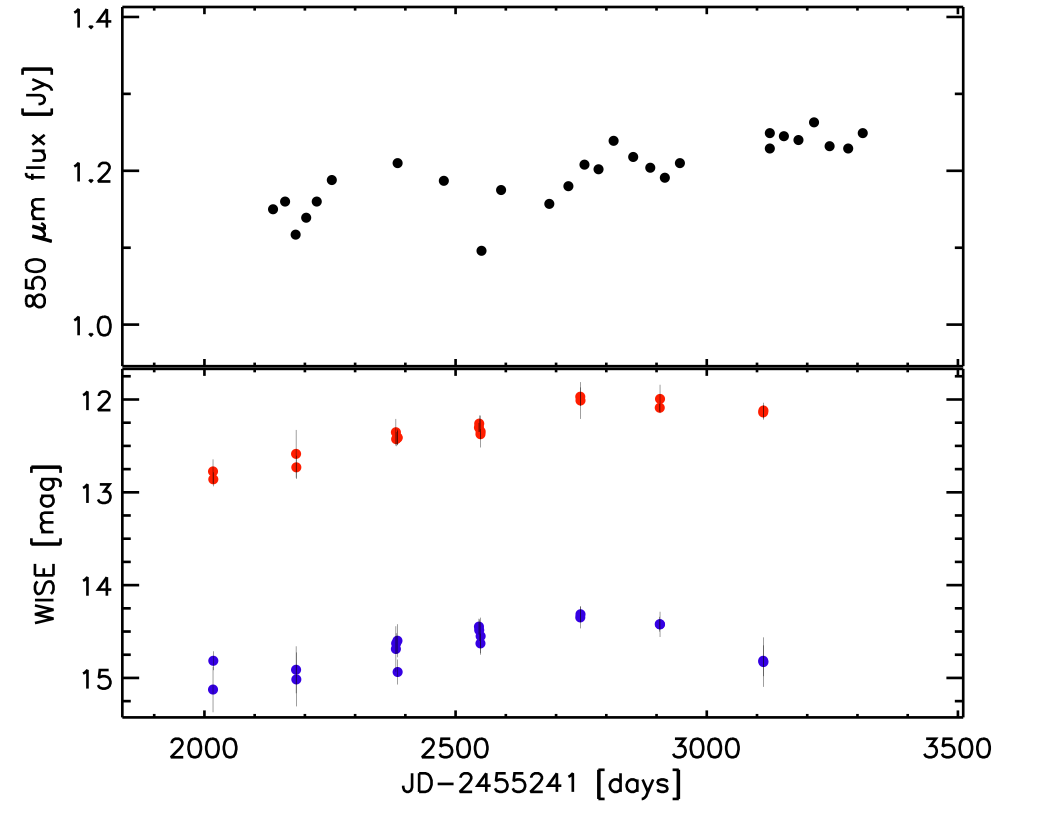}
  \includegraphics[width=0.45\textwidth]{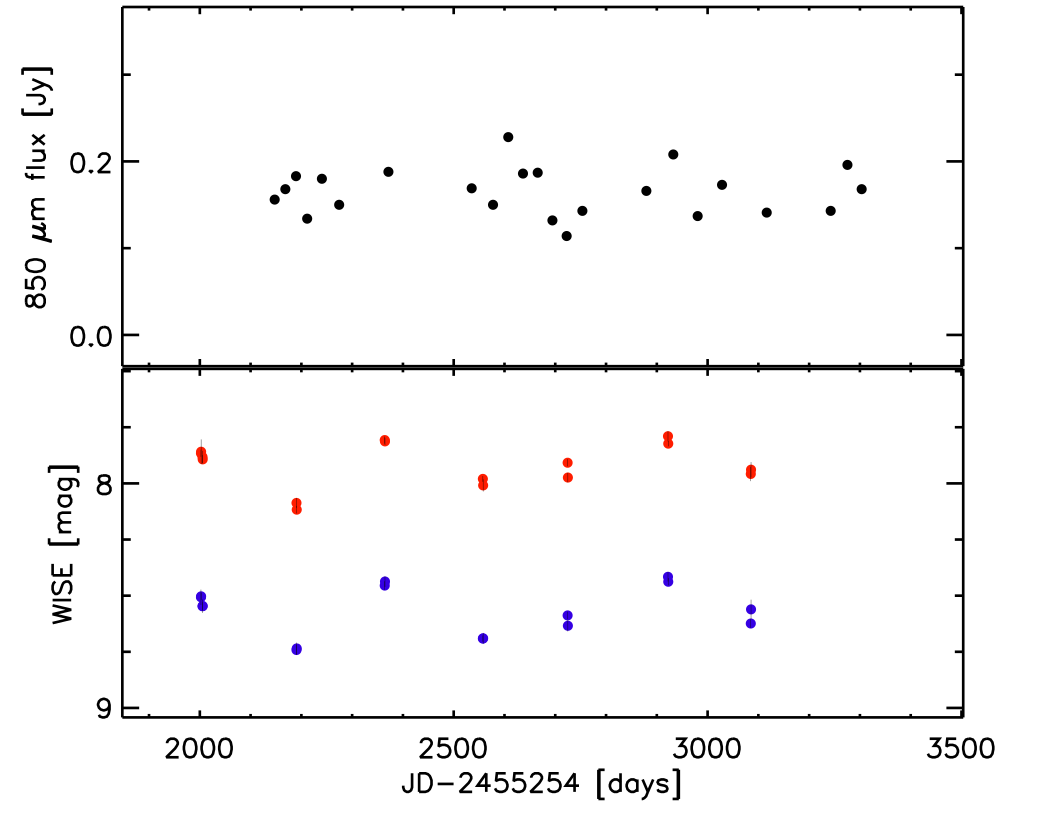}
  \caption{Preliminary figures from Contreras, et al. in prep. \textit{Left:} A confirmed long-term variable in the IC348 field at 3.4 (blue), 4.6 (red), and \mbox{850 $\mu$m} (black). \textit{Right:} A stochastic variable candidate in the Ophiuchus Core field. The colour scheme is the same as the left panel.}
  \label{fig:IRNeoWise}
\end{figure*}


In the case of the infrared wavelengths, if the emission can escape the envelope, the expected signal 
from an accretion burst should be much more significant as the YSO luminosity is being traced directly rather than tracing the temperature of the 
surrounding dust.  However, near- and even mid-IR variability may also be caused by changes in extinction (for example, V1647 Ori), while far-IR and submillimetre variability may be caused only by changes in luminosity.
This mid-infrared data (3.4 and \mbox{4.6 $\mu$m}) is available toward the Transient Survey fields throughout much 
of the time the JCMT has been obtaining images with \textit{WISE} and \textit{NEOWISE} (Contreras, et al. in prep). 
In the left panel of Figure 
\ref{fig:IRNeoWise}, we see a long-term brightening trend in both the MIR and Submillimetre data of an embedded YSO in the IC348 
region. 
In the right panel of Figure \ref{fig:IRNeoWise}, we see an example of a stochastic variable candidate in the Ophiuchus Core region with peaks and troughs in the mid-IR but not the submillimetre photometry. 
Using these observations as constraints, 3D and simplified 2D hydrodynamic modelling plus radiative transfer of protostellar variability has been developed
to interpret the SED variability of generic variables \citep{macfarlane19a,macfarlane19b} and for EC 53 specifically (Baek et al. in prep).  These models are needed to convert the submillimetre variability into a change in source luminosity while also allowing us to investigate the envelope structure, including outflow cavities and viewing inclination.  

The JCMT Transient Survey has also motivated several ALMA programs to resolve the changes in flux at high spatial resolution. The calibration methods developed throughout this single-dish program are now being applied in pilot ALMA surveys to understand whether it is possible to apply the same techniques to interferometric data \citep[e.g.][]{Francis2019}. 



\section{The Next Decade}
\label{sec:future}

\subsection{The Near Future: SCUBA-2}

The JCMT Transient Survey has definitively shown that variability associated with young stellar objects can be identified and characterised at submillimetre wavelengths. The current observing strategy of a monthly cadence toward 8 nearby star-forming regions has proven to be successful. Therefore, similar observations will be continued in the next generation programs in order to extend the 4-year timeframe, to better quantify any underlying timescales through periodogram analyses, and to construct deeper maps of fields densely populated with YSOs. The benefit of longer timescales was shown by \citep{Mairs2017GBS}, where Transient Survey data was compared with Gould Belt Survey data taken 2-4 years previous. A slow, long-term change in brightness can only be detected and verified over many years. The longer the timescale, the more sensitive the analysis is to identifying these ``secular'' variables. Additionally, with 3-4 more years of data collected on these same regions, several epochs obtained close in time can be combined in chunks to reduce the RMS background noise. By sacrificing some time resolution in this manner, fainter sources with long-term trends would be tracked with more certainty.

While the baseline large program will remain the same in principle, there are several ways to improve the observations in the near future. In regions that have the highest density of YSOs, a higher cadence of 1-2 weeks can be adopted. This increase in cadence will double the observing time for these fields with the current technology, but allow for the detection of shorter-period variability modes for bright sources while providing a factor of  $\sqrt{2}$ decrease in RMS noise relative to the current large program over monthly timescales. The increase in sensitivity over one month would allow for a significantly more robust calibration of $\sim$25\% more protostellar sources than are currently being tracked (see Section \ref{sec:NewCamera} for more details). 

While observing at a higher cadence can increase the monthly sensitivity of observed fields, targeting additional fields also bolsters the amount of sources observed and improves statistics. Further to the 8 fields that are currently being observed by the JCMT Transient Survey, there are 5-8 other regions that were observed by the Gould Belt Survey that have a high density of compact sources associated with known YSOs (this is necessary for relative flux calibration). These regions span Southern Orion A \citep{Mairs2016}, The W40 complex \citep{Rumble2016}, and \mbox{IC 5146} \citep{Johnstone2017}. The benefit of targeting these regions is that they can be compared to observations taken before 2016 in order to investigate long-term secular variability. An additional 8 regions would double the amount of observing time required for the survey. 

There is also significant interest from the community to expand the scope of the survey to regular observations of intermediate and high-mass star forming regions such as \mbox{NGC 2264} \citep[e.g.][]{Peretto2006}, \mbox{IC 3196} \citep[e.g.][]{Sicilia-Aguilar2014}, or the nearby Planck Galactic Cold Clumps already observed by the JCMT throughout the TOP-SCOPE large program \citep{Liu2018}. The further distances of these regions ensure the observation of more protostars per unit area at the cost of source confusion within the field. High-mass protostars likely undergo more energetic events, and the presence of high-mass protostars means that the field will contain many more low-mass protostars to build the survey's statistics. An initial analysis of SCUBA-2 observations towards 12 TOP-SCOPE fields \citep{Park2019arXiv} identified one candidate variability event in a high-mass star-forming region, but the analysis was limited because only the region was observed only three times. Improvements to the relative flux calibration pipeline would need to be made to ensure consistently measured light curves for these further regions. 

Outside of the main large program that will continue to monitor variability over selected regions, there are several opportunities for complementary PI projects. For example, Target of Opportunity (ToO) time will be vital if another survey, (e.g. ZTF or Gaia Alerts) announces the detection of an event at a different wavelength (ToOs were triggered using ALMA and IGRINS as a result of the JCMT observations of EC 53), or, if simultaneous observations involving multiple facilities are to be coordinated. The relationship between the JCMT and The Submillimetre Array (SMA) will be invaluable during these times, observing the same event from the same physical location. Interferometers such as the SMA and the Atacama Large Millimetre/submillimetre Array (ALMA) offer resolution and sensitivity to observe small fluctuations in brightness at the scale of the disk where
episodic accretion may be driven. Recently, \citep{Francis2019} presented novel methods for comparing time-series interferometric observations using Combined Array for Research in Millimeter-wave Astronomy (CARMA) and ALMA 1.3mm observations of deeply embedded protostars in Serpens taken 9 years apart. High resolution spectroscopic follow-ups of variability events at facilities such as Gemini, Keck, and SOFIA are also being pursued to evaluate the physical cause of the instability by evaluating inner disk heating and binarity.

\begin{figure}
  \centering
  \includegraphics[width=0.45\textwidth]{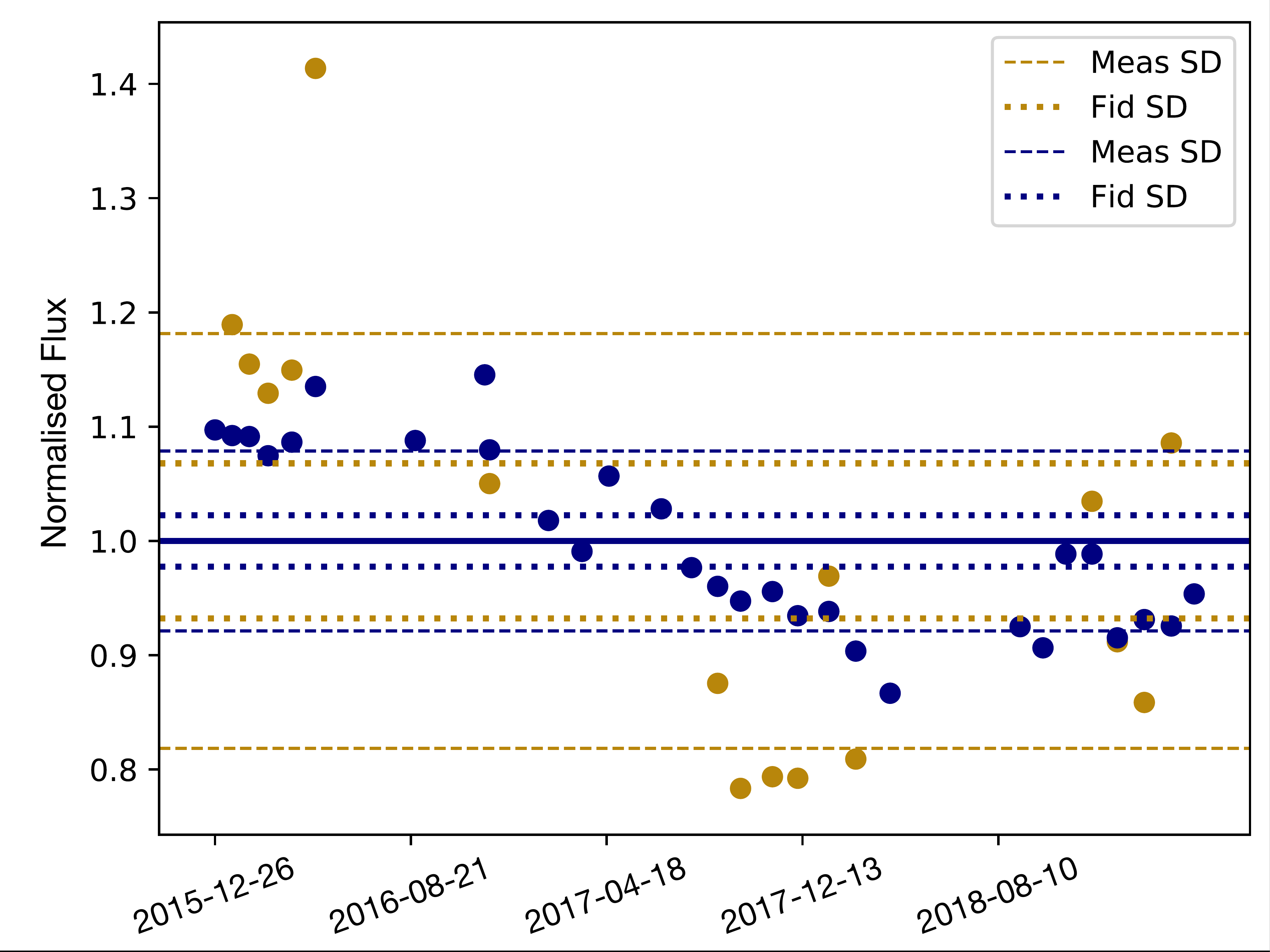}
   \includegraphics[width=0.45\textwidth]{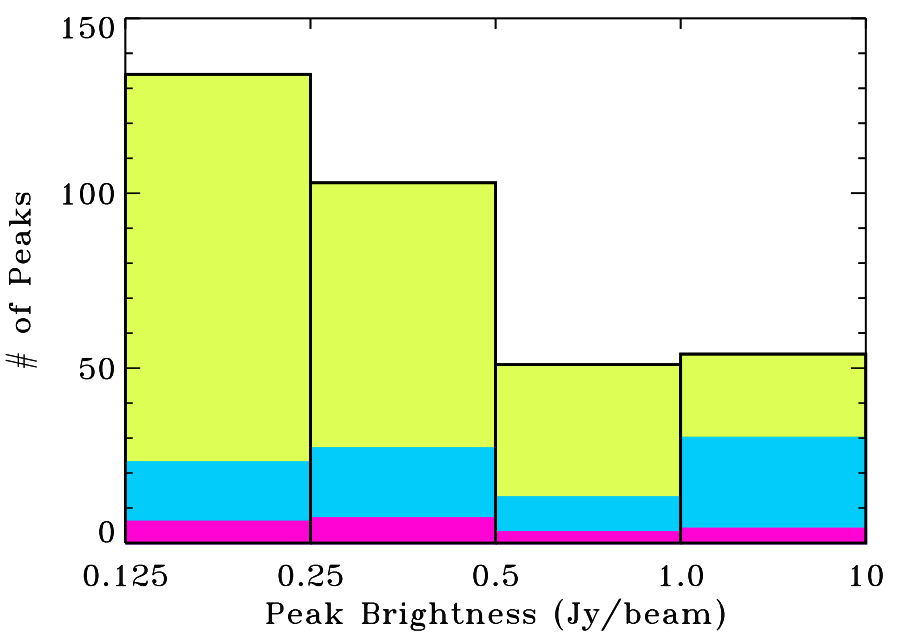}
  \caption{{\it Left:} HOPS 358 light curve shown at 450 (gold) and \mbox{850 $\mu$m} (blue). Dashed lines show the measured standard deviation in each light curve while dotted lines show the expected standard deviation. {\it Right: }From \protect\citep{Herczeg2017}. Distribution of 342 sources that have compact \mbox{850 $\mu$m} emission with peak brightness above
0.125 Jy/beam for all eight regions in our survey (yellow). The purple and blue
histograms, respectively, show the number of sources associated with one or
more disks and protostars. Based on the analysis of \protect\citep{Mairs2017Cal}, with the current SCUBA-2 set-up, we
can achieve 2--3\% accuracy for the 105 peaks brighter than 0.5 Jy/beam and
10\% for the 237 sources with brightness 0.125-0.5 Jy/beam.}
  \label{fig:HOPS358-450850}
\end{figure}


\subsection{Beyond SCUBA-2}
\label{sec:NewCamera}

Despite the successes of the JCMT Transient Survey so far, the current program is limited by statistics due to the depth of each observation and the small number of observed fields. Figure \ref{fig:HOPS358-450850} shows a histogram of more than 300 \mbox{850 $\mu$m} compact emission sources with peak brightnesses above \mbox{0.125 Jy/beam} identified over all 8 survey regions \citep{Herczeg2017}. The purple and blue histograms show  the fraction of those sources associated with one or more known Class II (disk) object or Class 0/I/Flat spectrum protostar, respectively. The typical RMS noise in a given \mbox{850 $\mu$m} Transient Survey image is approximately 0.015 Jy/beam. For sources that have a peak brightness greater 
than \mbox{0.5 Jy/beam}, the relative calibration uncertainty 
is $\sim2-3\%$. There are 105 emission sources detected in this brightness range, 42 of which are known to be 
protostellar. Expanding this high-accuracy bracket to include all sources brighter than \mbox{0.125 Jy/beam} would allow for the robust tracking of 
an additional 237 emission sources, 51 of which are known to be protostellar. Achieving this accuracy with SCUBA-2 requires a background RMS noise of $\sim 0.005$ Jy/beam, 3 times fainter than the current value. This factor of $\sim10$ increase in observing time required is prohibitive with SCUBA-2 but achievable with the proposed \mbox{850 $\mu$m} MKID camera.

The new JCMT instrument is expected to increase the mapping speed by a factor of 10 from the combination of more sensitive detectors and a wider field of view. As shown in Figure \ref{fig:HOPS358-450850}, this would dramatically 
increase the number of monitored young stellar objects, bolstering the ability to observe robust variability events for the same amount of observing time ($\sim$50 hours per year). In the first 3 years of using the new \mbox{850 $\mu$m} camera, we expect more than a factor of 3 increase in variability detections in the regions that are currently monitored by both increasing the number of sources we can analyse at $\sim 2$\% precision and by uncovering smaller-scale variability on the bright objects in our current survey. Fainter events are expected to occur more frequently than their brighter counterparts. More short-timescale flares such as \mbox{JW 566} \cite{Mairs2019} will also be detected.
 In addition to tracking the flux variability of known YSOs, detecting significant variable flux associated with identified ``starless cores'' could lead to the discovery of first hydrostatic cores or deeply 
embedded protostars that were previously missed. 
Circular fields of 0$^{\circ}$.5 diameter are ideal for monitoring YSO variability at a range of distances. Therefore, gaining an increased mapping speed with respect to SCUBA-2 solely by covering a larger field of view to the same depth will not be of benefit to the goals of YSO transient science, but may benefit other projects.  

Future generation surveys with the new camera would uncover variability in more clouds spanning a variety of galactic environments. Wider area coverage is essential in order to dramatically increase the sample size of observed YSOs while allowing for cross-comparisons of light curves for stars forming in different physical conditions. With SCUBA-2 technology, the projects and survey expansions previously discussed in Section \ref{sec:future} require significant increases in observing time to achieve a fraction of what would be possible with the new camera assuming time investments comparable to what is currently spent on variability studies. As an example, the entire JCMT Gould Belt Survey area (103 0$^{\circ}$.5 diameter circular fields) could be monitored monthly to a depth of 0.015 Jy/beam for an investment of only 10 hours per year more than the current survey (a total of 60 hours per year, or 5 observing nights, in Band 3 weather), which only covers 8 such fields. Alternatively, a full repeat of these fields to a depth of 0.005 Jy/beam (similar to the original survey and three times deeper than a single scan obtained by the current JCMT transient survey) could be performed in the same amount of time. This single epoch of all survey fields at such a deep sensitivity would provide critical flux information on thousands of compact sources when compared to the data obtained in $\sim$2015.  

The MKID detectors in the new camera will be much more stable than the current TES detectors in SCUBA-2, based on tests performed for similar arrays \citep{Lourie2018,Bryan2018}. This stability is expected to translate into a lower uncertainty in the relative flux calibration, improving the confidence with which more distant and higher mass star-forming regions can be measured. In 30 hours of Band 3/4 observing time with the MKID detectors, all the fields in the SCOPE survey \citep{Liu2018} could be revisited. Additionally, there is a wealth of star-forming regions previously observed by the JCMT such as M17, DR 21, and S255\footnote{See the JCMT Archive: \url{https://www.cadc-ccda.hia-iha.nrc-cnrc.gc.ca/en/jcmt/}} that have great potential in showcasing variability associated with intermediate and high-mass forming stars. 

Characterising the frequency and amplitude of submillimetre variability events through a range of masses and galactic environments over several year timescales will be essential in determining a complete theory of star formation. The constant monitoring of regions over a several 
year timescale would also result in the deepest submillimetre maps ever obtained of these regions, creating many opportunities for ancillary science. 
Expanding the scope of this new field of time domain science with a factor of 10 increase in observing efficiency will ensure new discoveries and the initiation of related studies for many years. These unique insights into the nature and evolution of the youngest stars will only be possible with the new \mbox{850 $\mu$m} camera at the JCMT.  


\subsection{A Note Concerning 450 $\mu$m Data}

Throughout all of these future prospects, JCMT Transient science will benefit greatly from the \mbox{450 $\mu$m} data that is collected simultaneously with the \mbox{850 $\mu$m} data. While the main focus of the Transient monitoring has been based on the \mbox{850 $\mu$m} data, 
there is ongoing work to make full use of the simultaneous \mbox{450 $\mu$m} data and to compare light curves of more evolved, less embedded 
YSOs across several mid to near-infrared filters. The relative flux calibration at \mbox{450 $\mu$m} is more challenging due to the higher impact of atmospheric water vapour on the quality of the data. Despite this, preliminary studies have shown that >60\% of the data can be recovered and flux calibrated to an uncertainty of 4-6\% (Mairs et al. In Prep). The simultaneity of this data is paramount in studying variability events at these wavelengths, especially when considering short-term, non-thermal events such as JW 566's stellar flare \citep{Mairs2019}. While the \mbox{850 $\mu$m} data can detect and reliably trace variability over time, additional wavelengths are necessary to constrain the physical conditions that are responsible for the event. So far, very good agreement is observed when comparing the 450 and \mbox{850 $\mu$m} light curves of known variable sources (see Figure \ref{fig:HOPS358-450850} for an example). Studies on the scaling relations, spectral indices, and phase shifts of these sources are ongoing, though preliminary efforts show stronger variations at 450 $\mu$m than at \mbox{850 $\mu$m}, consistent with theoretical expectations.

\subsection{Other science cases for submillimetre monitoring}
\label{sec:otherscience}

Our focus in this white paper and in our initial application for submillimetre monitoring has been variable protostars.  In contrast, optical monitoring surveys are designed to find supernovae, with applications for novae as well as more exotic phenomenon, like tidal disruption events and mergers of neutron stars \citep[e.g.][]{holoien17,graham19}.  Since dust obscures some transient events at optical wavelengths, ground- and space-based monitoring surveys are starting to monitor smaller regions on the sky in the near and mid-IR \citep{contreras17,kasliwal17}.  The detection of distant variable phenomenon will always be challenging for the JCMT because of the beam size and the sparse density of galaxies on the sky.  However, a new and more sensitive detector would open unique and powerful capabilities for monitoring eruptive phenomenon beyond nearby star-forming regions, including the distant universe and nearby galaxies, such as M31 and M33. Several specific science cases are described below. 

{\it Supernovae:} Some core-collapse supernovae in starburst galaxies may be hidden from the view of optical observers but would be detected in the mid-IR and submillimetre, thereby providing a more accurate measurement of the star formation and supernova rates in the most active star-forming galaxies \citep{yan18}.  Submillimetre monitoring of supernovae would also provide direct measurements of the feedback of supernovae by dust heating, and non-thermal emission as the supernova shock wave interacts with the surrounding interstellar medium.

\begin{figure}
  \centering
  \hspace{-5mm}
  \includegraphics[width=0.4\textwidth]{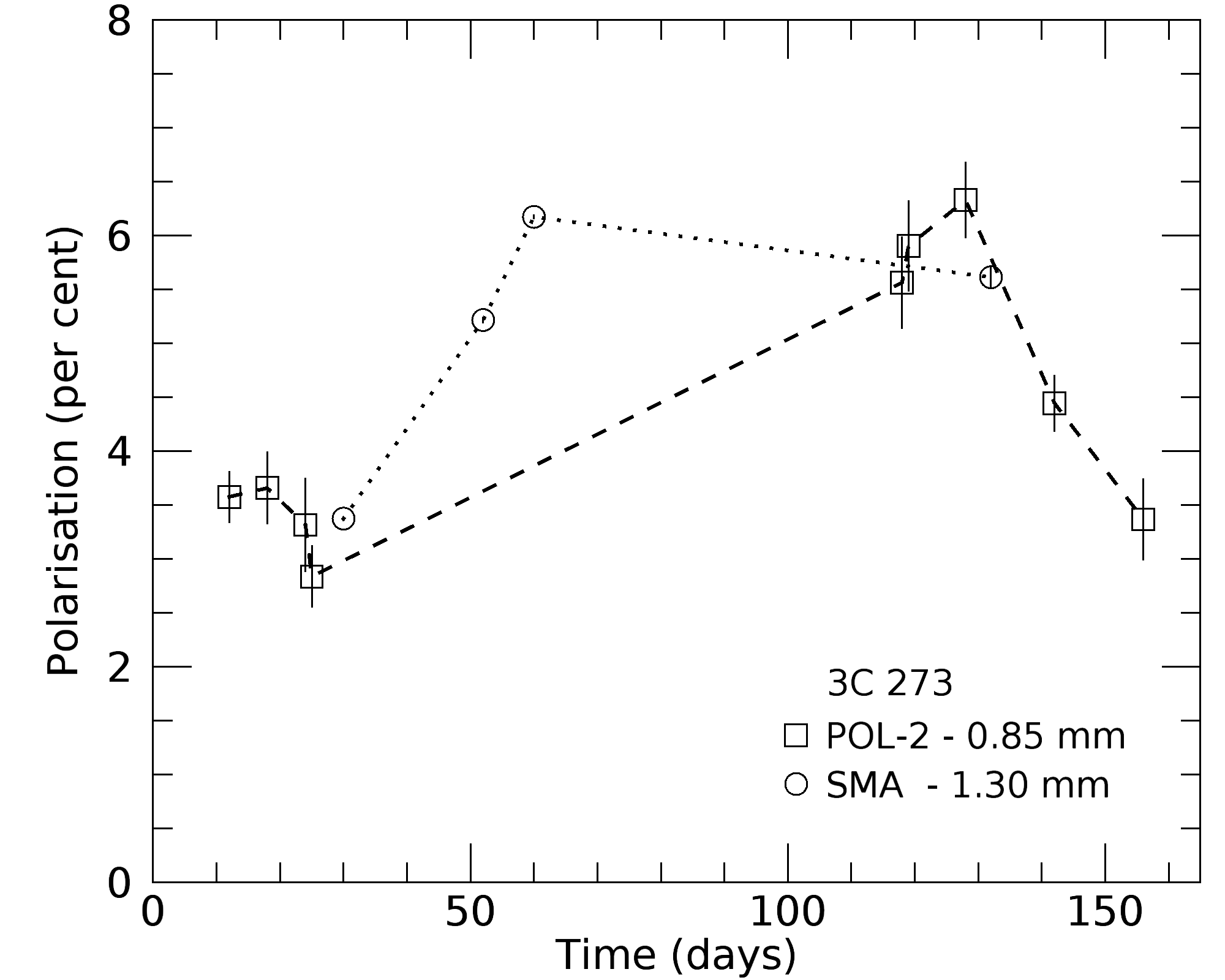}
  \hspace{5mm}
    \includegraphics[width=0.4\textwidth]{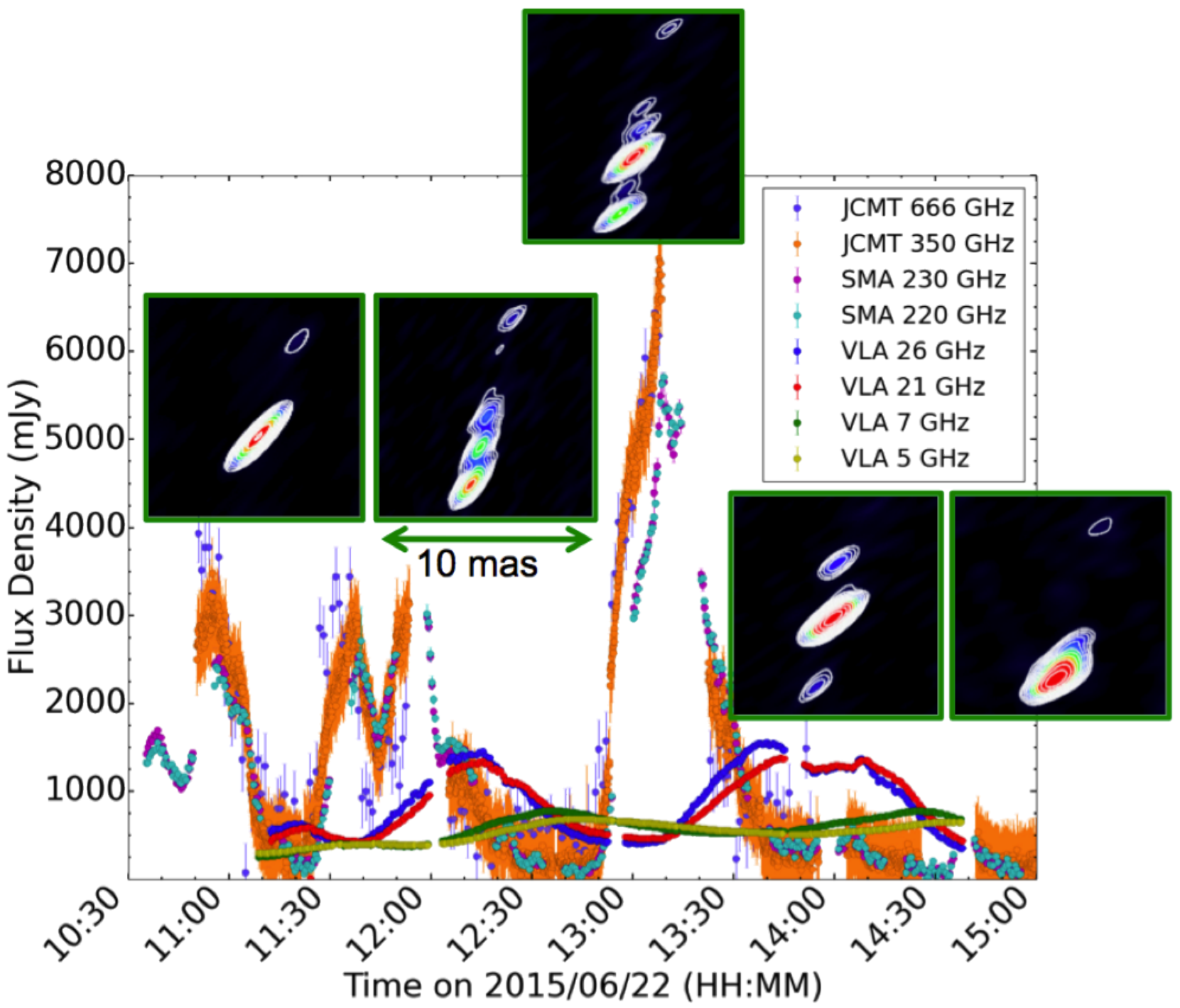}
  \caption{{\it Left:} Temporal variability (in days) of the polarization towards 3C~273 as a percentage of the continuum intensity at 850~$\mu$m (POL-2, shown as squares) and 1.3~mm (SMA, shown as circles). These data were obtained in 2016 by the POL-2 commissioning team as part of a polarimetric monitoring campaign of radio-loud active galactic nuclei.  {\it Right:} Time resolved light curves of the jet launched from the stellar mass black hole V404 Cygni, taken simultaneously in an unprecedented 8 different bands \citep{tetarenko17}. We detect rapid flux variability, in the form of multiple, large scale flaring events, which coincide with the launching of discrete jet ejecta (shown in inset panels of high angular resolution images taken with the Very Long Baseline Array; \citep{mj19}). This work represents the first time-resolved submillimetre study of XRB jets, and demonstrates the paramount importance of the submillimetre bands in understanding the rapidly evolving jets in XRB systems.}
  \label{fig:3c273}
\end{figure}


{\it Active Galactic Nuclei:} Synchrotron radiation from radio-loud active galactic nuclei (AGN) produces variable non-thermal emission \citep[e.g.][]{Robson2001} that probes the innermost components of relativistic jets, which are launched by accretion events onto supermassive black holes \citep[e.g.][]{Marscher2006}. The highly turbulent nature of the magnetized medium found within shocks along these relativistic jets may be responsible for their observed temporal variability in both continuum and polarised intensity \citep[e.g.][]{Jorstad2007, Marscher2014}. Such a variability in polarisation has successfully been observed at 850~$\mu$m on a timescale of days for the blazar 3C~273 using the POL-2 polarimeter at the JCMT (see Figure~\ref{fig:3c273}). However, there is also evidence for intraday variability in the polarization of blazars at millimetre wavelengths \citep{Lee2016day}. Similar flaring has been detected from Sgr A* at our galactic center \citep{yusef09}.  Even larger variability attributed to tidal disruption events may be deeply obscured by dusty torus.  A submillimetre camera capable of sensitive, high-cadence polarimetric measurements would therefore be an invaluable asset to probe the shortest coherence timescales of magnetized turbulence in AGN jets and other related variability of supermassive black holes.

{\it X-ray binaries:}  Stellar mass black holes  existing in X-ray binary (XRB) systems in our own Galaxy (i.e., binary systems containing a black hole accreting matter from a companion star) also launch highly variable relativistic synchrotron jets \citep{fend06}. When compared to AGN, the rapid evolutionary timescales of XRBs (days to months, rather than $10^6$ yrs), offers a distinct observational advantage, allowing us to watch the jet and accretion flow change on real human timescales.
While XRB jets emit across a wide range of frequencies, the submillimetre bands uniquely probe the jet base region, where the jets are first launched and accelerated. 
Detecting and characterizing rapid flux variability in jet emission from multiple XRBs can allow us to track accreting matter from inflow to outflow, and probe detailed jet properties that are difficult, if not impossible, to measure by traditional spectral and imaging methods (e.g., size scales, speed, the sequence of events leading to jet launching; \citep[e.g.][]{cas10,utt15,vinc18,mal18}). While XRB jet variability studies have been mainly confined to the higher frequency bands (optical, infrared), recently Tetarenko et al. \citep{tetarenko17,tet19} have extended these time domain studies into the radio and submillimetre bands, utilizing sophisticated Bayesian modelling and time domain techniques (e.g., cross correlation analyses, Fourier analyses) to directly connect jet variability properties to internal jet physics (see Figure~\ref{fig:3c273}). However, with current instrumentation (e.g., SCUBA-2 of JCMT), these studies could only probe jet variability in the brightest XRB systems, showing hundreds of mJy to Jy flux density levels. A more sensitive submillimetre camera would allow us to sample fainter, more common XRB jets, and significantly probe variability over much shorter timescales.

{\it Evolved Stars:}  The pulsations of AGB and other evolved stars may affect their dust shells, leading to variable continuum emission \citep{planck2015}.  submillimetre monitoring of evolved stars presents a number of advantages over the traditional optical/NIR observations. Not only is the submillimetre free from extinction, and avoids confusion caused by changes in spectral type by primarily detecting the circumstellar emission, but it has the potential to directly probe the influence of the variations on the outflow, rather than having to infer them indirectly: radio photospheres, dust and molecules all contribute to the submillimetre emission. By studying variability in the submillimetre and relating that to the behaviour of the stars themselves (as probed in the optical and NIR) we can unravel the influence of the pulsations on the inner envelope, where the outflow is launched. This is particularly interesting for the most optically-thick (i.e. highest mass-loss rate) sources, where even the mid-infrared is obscured, and for supernova progenitors, where it may shed light on the mechanisms driving pre-supernova mass loss, the most important unknown in the ultimate evolution of massive stars. Typically, one hour of integration spread over several epochs should be sufficient to smoothly sample the light curve for most nearby sources.

{\it Flare stars:}  Non-thermal radio flares from stellar reconnection may also be a potential area of expansion for JCMT, with feasibility demonstrated by the detection of a flare from JW 566 \citep{Mairs2019}.  Most observations of radio flares have focused on longer wavelengths, however coordinated observations that include the submillimetre and radio wavelengths would lead to a spectral index that would correspond to the electron opacity, while monitoring simultaneous to X-ray emission reveals connections between electron acceleration in magnetic fields and the production of high-energy photons \citep{forbrich2017}.

\subsection{JCMT in a Global Context}

A new \mbox{850 $\mu$m} camera at the JCMT would be uniquely capable of carrying out effective, 
consistent submillimetre Transient science observations due to its fast mapping speed in 
combination with a relatively wide field of view. The physical location of the JCMT in Hawai`i provides the capability of covering both northern and some southern fields. All this leads to plentiful synergies with other facilities in the coming decade. For example, pairing JCMT data with future monitoring campaigns undertaken by the Large Millimetre Telescope (LMT), the Institute for Radio Astronomy in the 
Millimeter Range (IRAM) 30-m dish, and near-future facilities such as the James Webb Space Telescope (JWST) and the Cornell Caltech Atacama Telescope Prime (CCAT-p). Follow-up studies on JCMT-discovered variable sources with the Atacama Large Millimetre/Submillimetre array (ALMA) will also be highly beneficial for probing the characteristics of the youngest stars. Since the JCMT will remain the most stable and reliable facility for \mbox{850 $\mu$m} monitoring, the potential for new and continuing collaborations is strong world-wide. JCMT Transient science will drive the development of innovative programmatic infrastructure to combine data at multiple wavelengths over rapid timescales, ushering in next-generation instrumentation and ambitious science goals with the space telescopes of the 2030s.

%
%

\bibliographystyle{unsrtMaxAuth} 
\bibliography{references}  

%
%
%

\end{document}